# Imaging of Strain Driven Magnetic Domains and Strong Spin-Phonon Coupling in Epitaxial Thin Films of SrRuO$_3$

Shekhar Tyagi[1], V.G. Sathe[1], Gaurav Sharma[1], Rajeev Rawat[1], Mael Guennou[2], Jens Kreisel[2]

[1]UGC-DAE Consortium for Scientific Research, University Campus, Khandwa Road, Indore-452001 India

[2]Physics and Materials Science Research Unit, University of Luxembourg, 41 Rue du Brill, L-4422 Belvaux, Luxembourg

**Abstract**

Epitaxial thin films of SrRuO$_3$ with large strain disorder were grown using pulsed laser deposition method which showed two distinct transition temperatures in Magnetic measurements. For the first time, we present visual evolution of magnetic domains across the two transitions using Magnetic force microscopy on these films. The study clearly showed that the magnetic anisotropy corresponding to the two transitions is different. It is observed that the perpendicular magnetic anisotropy is dominating in films which results in domain spin orientation preferably in out of plane direction. The Raman studies showed that the lattice is highly influenced by the magnetic order. The analysis of the phonon spectra around magnetic transition reveals the existence of strong spin-phonon coupling and the calculations resulted in spin-phonon coupling strength ($\lambda$) values of $\lambda \sim 5$ cm$^{-1}$ and $\lambda \sim 8.5$ cm$^{-1}$, for SrRuO$_3$ films grown on LSAT and SrTiO$_3$ single crystal substrates, respectively.

Functional oxides has been in attention and it is now realistic to envisage all-oxide electronic devices with the potential to compete with semiconductor architectures [1]. One of the fundamental requirement in many of the application where coupling among various degrees of freedom. SrRuO$_3$ has itinerant ferromagnetism and hence is a potential candidate for spintronic applications. The SrRuO$_3$ (SRO) has been under study for some time for oxide electronics for its electric and magnetic properties [2] such as uncommon transport properties, high-perpendicular remnant magnetization, large magneto-optical constant in thin films [3], bad metal behaviour at high temperatures [4], and Fermi liquid nature at low temperatures [5].The magnetic phase transitions shown by the bulk material around 160 K occurs at lower temperatures in thin films depending on the choice of the substrate and deposition conditions [6,7,8,9,10]. It appears that in this compound the spin is thus highly correlated with the lattice degree of freedom. However, the coupling constant of spin order with lattice is not reported so far in this compound. Similarly, the knowledge of magnetic domain structure and their corresponding magnetic anisotropy is missing. In our earlier study we have observed strong spin-orbit coupling in strain disordered SrRuO$_3$ thin films grown on LSAT and SrTiO$_3$ single crystal substrates. The strong spin-orbit coupling is shown to be the root cause for the relaxation of orbital quenching leading to the large magnetic moment shown by film grown on LSAT [11]. In another report [12] on La$_2$Co$_{0.8}$Mn$_{1.2}$O$_6$, Lopez-Mir et al had shown that strong spin-orbit coupling induces perpendicular magnetic anisotropy (PMA) which has a magneto crystalline origin.

However, the presence of uniaxial magneto-crystalline anisotropy due to the strong spin orbit coupling is unclear up to date in SrRuO$_3$. Previously reported results are contradictory about the orientation of easy axis in thin films and single crystals. For example, the single crystal single domain sample possesses its easy axis parallel to the [110] orientation [13] or [010] orientation [14]. The orientation of the easy anisotropic axis was reported to change towards the [001] axis in (Ca,Sr)RuO$_3$ single crystal with the increasing of Ca substitution [14]. The uniaxial perpendicular magnetic anisotropy in the [110] direction is reported for SRO thin films grown on SrTiO$_3$ (001) substrate, that is perpendicular to the film plane [15]. In another report, the authors observed the perpendicular anisotropy in the SRO thin films grown on STO (110) substrates [16]. Klein et al. reported that the angle between the easy axis and perpendicular axis of the film plane changes as a function of increasing temperature from 30$^0$ (at low temperatures) to 45$^0$ around magnetic transition temperature [4]. The strain relaxation and modification of strain due to buffer layer also affects the magnetic anisotropy in the films. For example, it was shown that the magnetic anisotropy is perpendicular in the case of SRO films grown on bare STO substrate, while it is in the plane of the film when buffer layer of Ba$_{1-x}$Sr$_x$TiO$_3$ is introduced between SRO and STO substrate [17]. Recently, Kolesnik et al. also observed that the magnetic easy axis lies in the [001] plane for the SrRuO$_3$ thin films while, it lies in between the [110] and [010] axes for the SrRu$_{0.92}$O$_3$ thin films and tilted away from the perpendicular to the surface by 23°-26° [18].

Thus, we feel that understanding landscape of the local strain distribution, magnetic anisotropy direction and coupling among spin and lattice degrees of freedom is key in understanding the magneto-elastic properties of this compound. A detailed study of the magnetic domains in this material is interesting from the point of view of fundamental physics as well as practical applications. Similarly, the study of spin-phonon constant is fundamental in exploration of this material in electronic devises. Magnetic force microscopy (MFM) has been widely used to map magnetic domain patterns and magnetic phase transitions on various materials [12,[19][20]]. In this article, the magnetic images of an epitaxial thin films of $SrRuO_3$ grown on LSAT (S1) and STO (S2) obtained using a variable-temperature MFM is presented. We have observed the evolution of domain patterns through the magnetic phase transition with perpendicular magnetic anisotropy. The temperature dependent Raman study showed remarkable spin-phonon coupling constant.

The thin films of $SrRuO_3$ grown on LSAT and STO (~185 nm thick) were thoroughly characterized using x-ray diffraction and reciprocal space map as reported previously [11]. The Raman spectroscopy was carried out using HR-800 Horiba Jobin Yvon, micro-Raman spectrometer equipped with a He-Ne laser ($\lambda$=632.8 nm). The spectral resolution of the system is ~1 cm$^{-1}$. Magnetic properties of the films was examined using 7-Tesla SQUID-vibrating sample magnetometer (SVSM; Quantum Design Inc., USA). Magnetic images were extracted using low temperature magnetic force microscope (MFM) from M/s. Attocube, Germany, along with a superconducting magnet system from M/s. American Magnetics, USA. Co coated n-Si cantilever (PP-MFPR from Nanosensor) with resonance frequency $\approx$ 75 kHz was used for the measurement and the applied magnetic field direction was kept perpendicular to the film plane. MFM images were measured in frequency shift mode with a constant lift of 50 nm and slope-correction option was used during scanning to care for average sample slope. Later was estimated from topography scan and these scan showed that the sample roughness is around 10nm. Detailed x-ray diffraction studies on SRO/ LSAT (S1) and SRO/ STO (S2) films confirmed epitaxial nature of the films and presence of strain disorder [11].

The bulk SRO is reported to show ferromagnetic order below $T_C$~160K with magnetic moment of ~ 1.2 $\mu_B$/Ru atom [21,22,23]. We have carried out the magnetization measurements as a function of temperature in the presence of 500 Oe applied magnetic field and the data was recorded in the field-cooled (FC) and zero-field-cooled (ZFC) protocol for the S1 and S2 films which is illustrated in figure 1(a) and 1(b), respectively. We found two magnetic anomalies; at ~ 153K (T1) and 135K (T2) in S1 film while at ~ 148K (T1) and 128K (T2) in S2 film. Second derivative of the FC cycle measurement shows discontinuous change in the slope across the magnetic anomalies that is illustrated in the inset of the figure 1(a) and 1(b). The two components are markedly different and thus, are likely to be related with the magnetic anisotropic states possibly induced due to strain disorder.

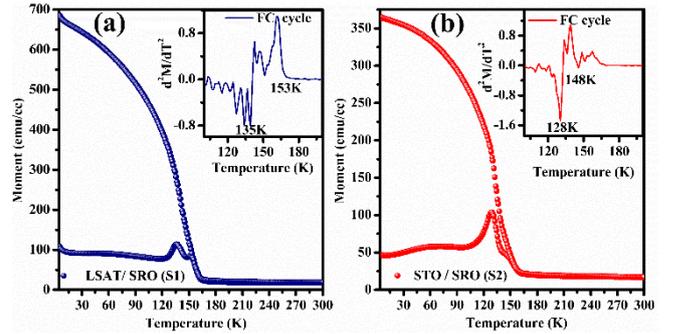

**Figure 1:** The magnetization (M) measurements as a function of temperature (T) carried out on LSAT/SRO (S1), (a) and STO/SRO (S2) thin films, (b) by using zero field cooled (ZFC) and field cooled (FC) protocol under 500 Oe field. Insets show the second derivative of the FC cycle across the magnetic transition temperatures.

MFM images are taken on the two films in the absence and presence of an external magnetic field (500 Oe) at three temperatures, below T2, in between T2 and T1 and above T1, in warming cycle. Magnetic force microscopy is an ideal tool to probe spatial evolution of such magnetic domains on sub-micron length scale. The images obtained by magnetic force microscopy on the two films are illustrated in figure 2 (a) and (b) corresponding to S1 and S2 films, respectively. All the images corresponds to 5 $\mu$m x 5 $\mu$m sample area. Figure 2 (a), shows four MFM images; recorded at 100 K under 0 Oe and 500 Oe field; at 145 K under 500 Oe field and at 170 K (paramagnetic phase) under 500 Oe field and their corresponding topographic images at the bottom. The magnetic images measured at 100 K under 0 Oe magnetic field shows prominent bright and dark regions representing well defined magnetic domains. In the present instrumental set-up, magnetic force with out-of-plane direction can only be detected, the dark (bright) regions represents magnetic domains with spin direction up (down) to cantilever magnetization/magnetic field direction, while the grey region represents magnetic domains away from it.

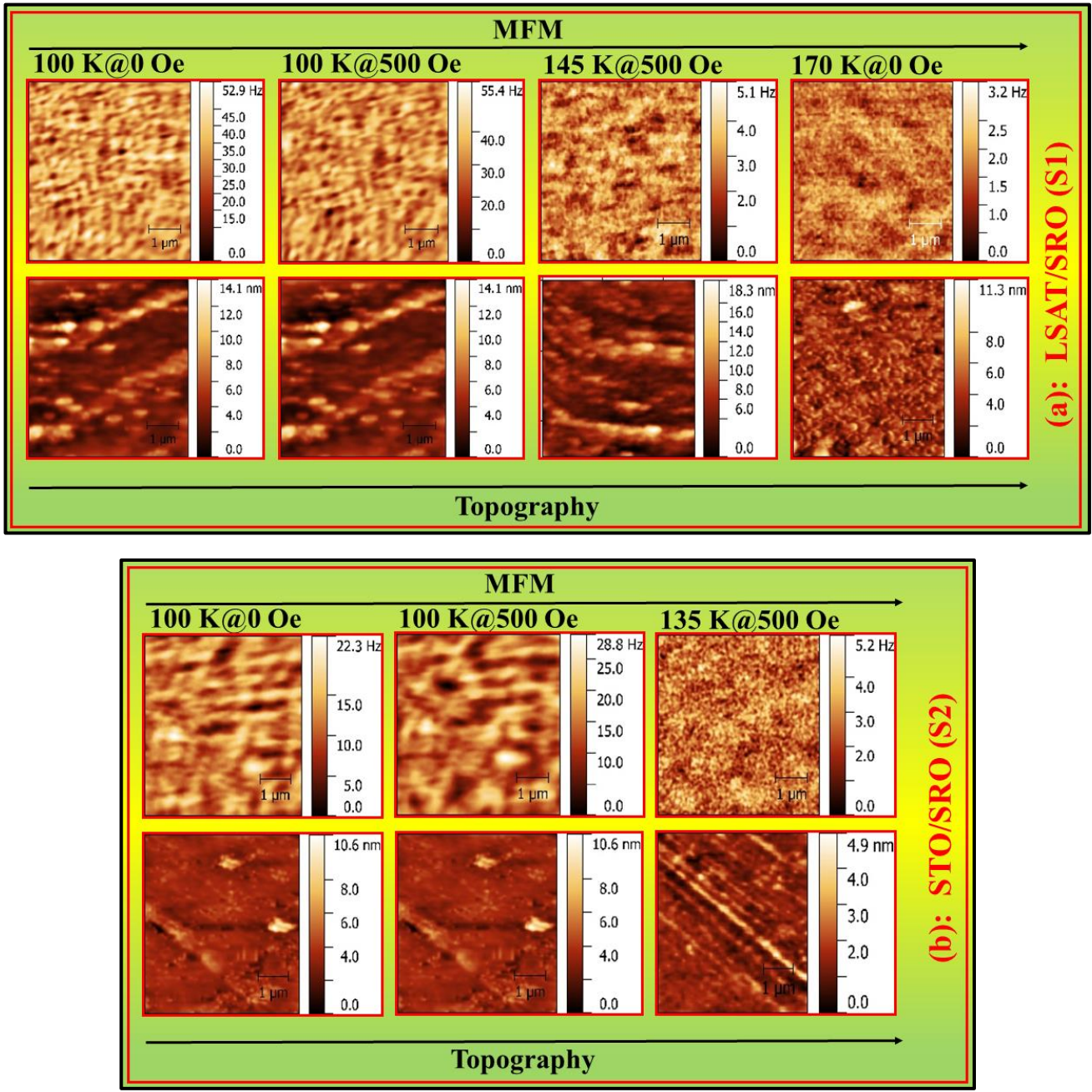

**Figure 2:** Magnetic Force Microscopic (MFM) measurements carried out as a function of temperature and applied external magnetic field; (a) on LSAT/SRO (S1) film (b) on STO/SRO (S2) film. The topographic images of the same area is presented at the bottom.

Therefore, at 100 K, it appears that the magnetization is lying in the out-of-plane as well as in the plane. The corresponding topographic image is also illustrated at the bottom for comparison. We applied a magnetic field (500 Oe) in the perpendicular direction of the film plane at 100 K and measured the MFM images at the same area and position of the films. The image remain nearly unaffected with the application of field. In the same applied magnetic field condition, the MFM images were recorded at 145K which is between T2 and T1. At this temperature, the contrast (frequency scale) was found to reduce by one order of magnitude, however, the dark and bright regions are still observable in the image. This observation suggests that the lower $T_C$ (T2) corresponds to the in-plane magnetic anisotropy (IMA) while the higher $T_C$ (T1) corresponds to the PMA. The images scanned at 170 K under the field of 500 Oe showed subtle contrast due to short range order typical of paramagnetic phase.

The measurements following the same protocol were carried out on film S2 and are shown in Figure 2 (b). MFM

images scanned at 100 K in the absence of applied magnetic field again showed amalgamation of bright and dark regions. The dark and bright regions are significantly larger in size in S2 when compared to that in S1. Further, in contrast to film S1, in this film the bright and dark regions are nearly equal in number and are well spread across the image. This confirms that at 100 K the spin direction is along out-of-plane direction of the film with statistical spread of up and down domains. On the application of the field, dark regions as well as the contrast improves. When the temperature is raised to 135 K in the presence of field the magnetic images shows diminishing magnetic contrast. This may be due to lower value of the magnetization in this film compared to magnetization value observed in S1.

In conclusion the MFM measurements gave direct evidence about mixed magnetic domains, different magnetic transition temperatures and magnetic anisotropy direction in the two films. It also provided a direct evidence of effect of strain disorder on the size of the domains. In our previous report, we have clearly shown that in film S1 the lateral strain disorder is significantly larger than in S2, this restricts the size of the domain in S1 in comparison to S2. This also provides a stronger pinning effect in S1 and hence on application of field the magnetic image remains unaffected.

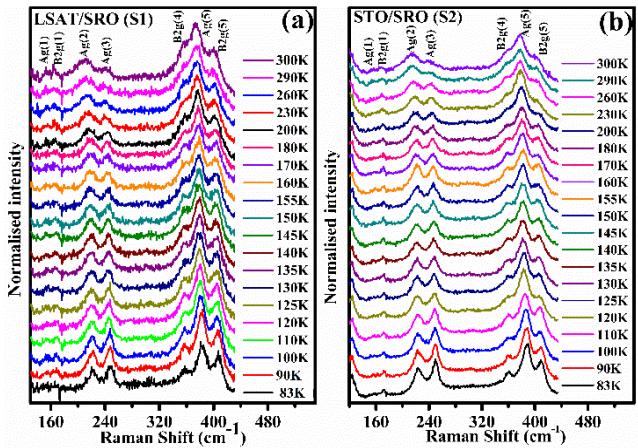

**Figure 3:** (a) and (b) The Raman spectra collected as a function of temperature for the LSAT/SRO (S1) and STO/SRO (S2) film, respectively.

The temperature dependent Raman spectroscopic study on S1 and S2 thin films is illustrated in figure 3 (a) and (b) respectively. Group-theoretical analysis for the Γ-point phonon modes of SRO (*Pnma* space group for *Z=4*) results in 60 Γ-point phonons out of which 24 are Raman-active [24]. However, we observed experimentally only 7 prominent Raman active modes in our films, Ag (1) at 123cm$^{-1}$, B2g (1) at 170 cm$^{-1}$, Ag (2) at 223 cm$^{-1}$, Ag (3) at 252 cm$^{-1}$, B2g (4) at 360 cm$^{-1}$, Ag (5) at 390 cm$^{-1}$ and B2g (5) at 410 cm$^{-1}$. The Raman spectra of the films matched with the previous report and we adopted the symmetries of the modes given in this report [24]. Noticeable changes are observed in the relative intensity of the Ag (3) mode and Ag (2) mode with decreasing temperature signifying increase in octahedral rotation.

The position and width of various Raman modes was obtained by fitting with Lorentz function that showed normal behaviour as a function of temperature barring B2g(4) mode. The Raman mode position of the B2g(4) as a function of temperature is plotted in figure 4 (a) and (c) for the sample S1 and S2 respectively. It showed dramatic changes around magnetic ordering temperatures. This mode represents apical oxygen vibrations which modulate the Ru-O-Ru bond angle. This induces Ru ion motion that modulates spin-exchange coupling [24 and references cited therein].

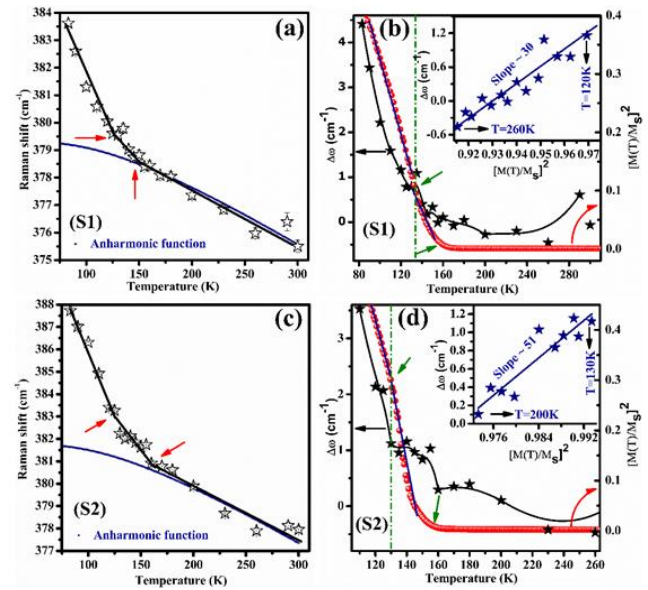

**Figure 4:** (a) and (c) The temperature dependence of B2g (4) mode wavenumber obtained by fitting the Raman spectra recorded on LSAT/SRO (S1) and STO/SRO (S2) film, respectively. The solid blue line show best fit using an-harmonic model (Equation 1). Arrow indicates the magnetic transition temperatures. Plot of $\Delta\omega$ and $[M(T)/M_s]^2$ with temperature for the B2g (4) mode corresponding to the S1 and S2 film is illustrated in (b) and (d), respectively. Solid (blue) line in the corresponding inset shows the linear least square fit to the data of $\Delta\omega$ versus $[M(T)/M_s]^2$ in the different temperature ranges. Vertical dashed line and arrows indicates the change in slope with temperature.

Generally, variation in the Raman mode position as a function of temperature is represented by an anharmonic model considering three phonon processes [25] as given below:

$$\omega(T) = \omega(0) - c\,(1+2/e^x-1) \quad \ldots\ldots\ldots\ldots\ldots (1)$$

Where, $x = h\omega_0/(4\pi k_B T)$; $\omega(0)$ and $\omega(T)$ are Raman frequency at temperature 0 and T K, respectively and c is the adjustable parameter.

Figure 4 (a) and 4 (c) shows the B2g(4) mode wavenumber as a function of temperature along with the calculated curve using best fit of equation 1 (solid blue line) to the experimental data above magnetic ordering temperature and extrapolated to the low temperatures for S1 and S2 film, respectively. The experimentally observed curve shows sharp variation from the theoretical curve below magnetic transition temperatures that is depicted by arrows.

The spin-phonon coupling strength was calculated using the method proposed by Granado et al. [26]. In this approach the exchange term is deduced using following equation,

$$<S_i \cdot S_j> = 6[M(T)/M_s]^2 \quad \ldots \ldots (2)$$

Where, $M(T)$ is the magnetization per Ru ion, $M_s$ is value of the saturation magnetization, and factor 6 appears due to presence of six nearest neighbour ferromagnetically coupled Ru ions. This suggests

$$\Delta\omega = (\lambda)<S_i \cdot S_j> = (\lambda) 6[M(T)/M_s]^2 \quad \ldots \ldots (3)$$

The $\Delta\omega(T)$ and $[M(T)/M_s]^2$ as a function of temperature for B2g(4) mode corresponding to S1 and S2 films is shown figure 4 (b) and 4 (d), respectively. A correlation between the frequency shift $\Delta\omega(T)$ and $[M(T)/M_s]^2$ is clearly evident in the temperature range of ~80 K to ~300 K, which suggests that the anomaly in the vicinity of magnetic transitions is due to the spin-phonon coupling in both the films. Vertical dotted line and arrows indicates the change in slope in magnetization with temperature. Solid (blue) line in the inset displays the linear least square best fit to the data between $\Delta\omega(T)$ versus $[M(T)/M_s]^2$ in the temperature range of 120 K to 260 K and 130 K to 200 K, for S1 and S2 film, respectively that was used to calculate the spin-phonon coupling strength $(\lambda)$ (1/6 of the slope). This resulted in a value of $\lambda \sim 5$ cm$^{-1}$ and $\lambda \sim 8.5$ cm$^{-1}$, corresponding to the S1 and S2 film, respectively. Our estimated strength of the spin phonon coupling is comparable to that in antiferromagnetic nickel oxide where it is 7.9 cm$^{-1}$ for TO and 14.1 cm$^{-1}$ for LO phonons [27] while it is larger than that in ZnCr$_2$O$_4$ ($\lambda \sim 3.2$ cm$^{-1}$, $-6.2$ cm$^{-1}$) [28]. Although, not as strong as in NaOsO$_3$ ($\lambda \sim 40$ cm$^{-1}$) [29] and CuO ($\lambda \sim 50$ cm$^{-1}$) [30] but our estimated values is perhaps the strongest for SrRuO$_3$ reported to date.

This study provides direct visual evidence of the magnetic domains of SrRuO$_3$ thin films across the magnetic transitions. The temperature dependent MFM studies revealed the magnetic anisotropy direction. At low temperatures, the films showed preference towards PMA. The study shows that the strain disorder arising during strain relaxation process plays important role in shaping domain sizes and anisotropy properties in SrRuO$_3$ thin films. The spin phonon coupling strength was estimated by Raman spectroscopy studies. The films showed remarkable spin-phonon coupling constant (between 5 cm$^{-1}$ and $\lambda \sim 8.5$ cm$^{-1}$) along with perpendicular magnetic anisotropy which makes this material suitable for applications.

We thank Dr. R. J. Choudhary for the magnetic measurements. We acknowledge Mr. Rohit Sharma and Mr. Sumon Karmakar for their help in MFM measurements.